%% file: main.tex
\documentclass[12pt]{iopart}

\usepackage{iopams}  
\usepackage{graphicx}

\newcommand{\pll}{{\parallel}}

\newcommand{\pd}{{\partial}}
\newcommand{\bm}{\boldsymbol}

\begin{document}

\title[Study on impurity hole plasmas by global neoclassical simulation]
{Study on impurity hole plasmas by global neoclassical simulation}

\author{Keiji Fujita$^1$ \
S. Satake$^{1,2}$ \
M. Nunami$^{1,2}$ \
J. M. Garc\'{i}a-Rega\~{n}a$^{3}$ \
J. L. Velasco$^{3}$ \
I. Calvo$^{3}$}

\address{$^1$ The Graduate University for Advanced Studies (SOKENDAI), 322-6 Oroshi-cho, Toki, Japan \\
$^2$ National Institute for Fusion Science 322-6 Oroshi-cho, Toki, Japan \\
$^3$ Laboratorio Nacional de Fusi\'{o}n, CIEMAT, Avenida Complutense, 40, 28040, Madrid, Spain
}
\ead{fujita.keiji@nifs,ac.jp}
\vspace{10pt}
\begin{indented}
\item[]May 2021
\end{indented}

\begin{abstract}
An impurity hole observed in the Large Helical Device (LHD) is a hollow density profile of an impurity ion species formed in the core plasma where the negative (inward-pointing) ambipolar radial electric field ($E_r$) exists.
Although local neoclassical models have predicted that the sign of $E_r$ in impurity hole plasmas is negative for the entire minor radius, an experimental measurement of an impurity hole plasma has shown that the $E_r$ changes the sign from negative to positive along the minor radius.
In the present work, we investigate neoclassical impurity transport in an impurity hole plasma using a global neoclassical simulation code FORTEC-3D.
The variation of electrostatic potential on each flux surface ($\Phi_1$) is evaluated from the quasi-neutrality condition in multi-ion-species plasma by the global simulation. 
The ambipolar $E_r$ and neoclassical fluxes are determined in solving a global drift-kinetic equation including the effect of $\Phi_1$.
By the global simulation, we show that an $E_r$ which changes the sign along the radius is obtained as a solution of the ambipolar condition and with such an $E_r$, impurity carbon flux can be outwardly directed even where $E_r<0$ and the carbon density profile is hollow around the magnetic axis.
Furthermore, it is found that the outward carbon flux is only a factor $2$-$3$ from balancing the modeled inward turbulent flux.
Our result indicates that we have moved one step closer to reproducing the impurity transport in impurity hole plasmas by kinetic simulation.
\end{abstract}

%
%
%
%
%

\section{Introduction}\label{sec:Intro}

\input{Body/Intro}

\section{Theoretical model}\label{sec:model}
\input{Body/DIR_MODEL/Equations}

\section{Set up for the calculation}\label{sec:setup}
\input{Body/DIR_SETUP/Species-and-EBfields}

\input{Body/DIR_SETUP/nTprof}

\section{Simulation result}\label{sec:result}

In this section, we present the simulation results of each case.
We focus on how and how much the global effects and the $\Phi_1$-effect give rise to the difference in the ambipolar condition and the radial particle fluxes. 
Some results of local neoclassical simulation without the $\Phi_1$-effect by PENTA code are also shown for reference.

\input{Body/DIR_RESULTS/ResultA}

\input{Body/DIR_RESULTS/ResultB}
\input{Body/DIR_RESULTS/ResultC}

\section{Summary and discussion}\label{sec:disc}
\input{Body/Discussion}

\input{Body/Acknowledgement}


\section*{References}

\bibliographystyle{unsrt}
\bibliography{main}

\end{document}

%% file: Body/Intro.tex
In addition to fuel deuterium or tritium, fusion plasmas contain a variety of ions including helium produced by fusion reactions or high-$Z$ ions such as tungsten or iron sputtered from the wall.  
The presence of such impurity ions in the core plasma can degrade the performance of the fusion reactor by diluting the fuel ions and radiating a significant amount of energy away.
Ions with higher charges cause those unfavourable effects more significantly.
Understanding the impurity transport process to prevent the impurity accumulation is thus a crucial task to realize the steady-state operation of fusion reactors.

In stellarator and heliotron type devices, the ambipolar radial electric field, $E_r$, can play a crucial role in determining the behavior of impurity ions. 
In fusion relevant plasma, in which the electron temperature $T_e$ and the ion temperature $T_i$ are comparable, the sign of the ambipolar $E_r$ usually becomes negative \cite{Ho1987, Maassberg1999}.
Ion species are thus driven inwardly by the negative $E_r$ and its impact is proportional to the charge $Z$.
In this aspect, tokamaks have an advantage over stellarators.
In tokamaks, neoclassical transport is intrinsically ambipolar and is independent of the ambipolar radial electric field.
A sufficiently large temperature gradient can thus drive the impurity ions out of the core.
This is called ``temperature screening'' \cite{Hirshman1976}.
Although it is shown that the contribution of the terms involving the $E_r$ to the radial impurity flux can be cancelled even in stellarators, the condition for the effect to take place is rather strict \cite{Helander2017,Newton2017,Calvo2018ste,Buller2018,Martin2020,Calvo2019}.

The prediction of impurity accumulation under the condition discussed above has been confirmed by experiments \cite{Burhenn2009}, but a few exceptional cases have been observed as well.
One of the exceptions is observation of the so-called``impurity holes" in the LHD \cite{Ida2009,Yoshinuma2009}, which indicates the formation of hollow impurity density profiles in the core.
Neoclassical transport simulations have predicted negative ambipolar $E_r$ and inward-pinch of carbon impurity for the impurity hole plasmas \cite{Ida2009,Ido2010,Mollen2018,Velasco2018}.
Gyrokinetic studies have also predicted inward impurity turbulent flux in impurity hole plasmas \cite{Mikkelsen2014,Nunami2016,Nunami2020}.
Therefore, there should be other mechanisms to explain the exhaust of impurity ions from the center of the plasmas.

The increase in the ion temperature and its gradient, which characterizes an impurity hole plasma, is thought to play an essential role in the impurity hole formation.
If the size of the negative $E_r$ is small and the temperature gradient is sufficiently large, it is shown that the exhaust of impurity ions by the temperature gradient is in fact possible even when  
the negative $E_r$ does contribute to the radial transport \cite{Velasco2018,Velasco2017}.
However, according to the analysis, the ion temperature should be even higher than observed to overcome the inward pinch of moderate-$Z$ (namely, $Z=6$ for carbon) and high-$Z$ impurities caused by the negative $E_r$ and the positive impurity density gradient.

In order to provide an explanation for the mechanism behind the impurity hole formation, a variety of attempts to extend the neoclassical analysis have been made.
In \cite{Nunami2017impacts}, the possibility that the ambipolar $E_r$ is shifted to be positive by an external torque and the radial carbon flux becomes outward is shown.
To change the sign of the $E_r$, 
however, a fairly stronger external torque than a tangential neutral beam injection (NBI) torque input in LHD experiments is expected to be required.

Another approach to extension is to consider the effect of the variation of electrostatic potential on each flux surface, which has recently been shown to have non-negligible impact on impurity transport \cite{Calvo2018ste,Mollen2018,Velasco2018,Garcia2013,Garcia2017}.
The variation of electrostatic potential is defined as the non-flux function part of the electrostatic potential, $\Phi_1\equiv \Phi-\Phi_0$, where $\Phi$ is the total electrostatic potential and $\Phi_0=\Phi_0(r)$ is the flux function part of $\Phi$.
Still, the neoclassical simulations for the impurity hole plasmas have shown that the $E_r$ is negative along the entire radius and the carbon flux is driven further inwardly by the effect of $\Phi_1$ \cite{Mollen2018,Velasco2018}.

The $E_r$ profile has been measured with a heavy ion beam probe (HIBP) in an LHD plasma in which an impurity hole formation is observed \cite{Ido2010}.
The measurement shows that the $E_r$ is indeed negative from the magnetic axis to $r/a=0.55$ as predicted by the neoclassical simulations, where $r$ is the radial coordinate and $a$ is the minor radius.
However, the same experimental data shows that the sign of the $E_r$ transits from negative to positive around $r/a= 0.55$ \cite{Ido2010,Nagaoka2011}.
The neoclassical predictions are thus, in fact, not accurate for $r/a>0.55$.
Still, it is mentioned in \cite{Velasco2017} that the condition of high-$T$ plasmas is close to that for $E_r$ to be positive.
This suggests a possibility that the observed profile can be numerically reproduced by improving the neoclassical simulation model.

The preceding numerical studies on the impurity neoclassical transport in the impurity hole plasmas have been carried out with the radially-local  drift-kinetic models, in which the magnetic drift velocity (grad-$B$ and curvature drift) of guiding-center motion is treated as a negligibly small quantity compared with the parallel and the $E\times B$ velocities. 
On the other hand, a global neoclassical transport code FORTEC-3D \cite{Satake2005,Satake2010,Matsuoka2011} has been developed to solve a drift-kinetic equation without relying on the radially-local approximation and taking the finite magnetic drift across the magnetic field line and the flux surface into account. 
Using FORTEC-3D, it has been revealed that the local approximation is inaccurate to evaluate the neoclassical transport in helical plasmas when $|E_r|$ is small and the $E\times B$-drift velocity is comparable to the magnetic drift velocity \cite{Matsuoka2015,Huang2017}, especially for ions and in non-neoclassical-optimized magnetic configuration such as in LHD. 

After extending the code for multi-particle species plasmas \cite{Satake2020}, the first applications of FORTEC-3D code to LHD impurity hole plasmas have been carried out in the trace-impurity limit, in which the ambipolar $E_r$ and $\Phi_1$ are assumed to be determined only from bulk hydrogen and electrons.
Through those applications, it has been found that the $\Phi_1$ potential profile on each flux surface obtained by the global calculation is different from that obtained with the local approximation models \cite{Fujita2019} and that the $E_r$ profile in the LHD impurity hole plasma is expected to be positive for almost the entire region \cite{Fujita2020}. 
It has also been found that the positive $E_r$ results in outward carbon impurity flux and the $\Phi_1$-effect further enhances the outward flux.
The global neoclassical simulation model thus becomes a candidate to explain the impurity hole phenomena found in LHD. 
However, there remain several issues to be considered.
First, the HIBP measurement detects the negative $E_r$ region in the center of an impurity hole plasma, which is inconsistent with the expectation from the FORTEC-3D simulation result. 
Second, the ambipolar $E_r$ and $\Phi_1$ have not been solved self-consistently in the trace-impurity limit.
In order to increase the reliability of calculations, we have thus enabled our code to simultaneously solve the quasi-neutrality condition to evaluate $\Phi_1$, the ambipolar condition to determine the ambipolar $E_r$, and a drift-kinetic equation for multiple ion species including $\Phi_1$.

In this article, we investigate the impurity hole plasma with the improved global code.
In the next section, we present the numerical model used in our calculation and show how we evaluate the radial particle flux, ambipolar radial electric field, and $\Phi_1$. 
In section \ref{sec:setup} the conditions of the plasma, such as the magnetic field configuration and the $n$-$T$ profiles, are shown.
In section \ref{sec:result}, the simulation results are shown and compared with previous studies.
The most striking outcomes of the global simulations are the ambipolar $E_r$ profile which is negative near the magnetic axis but transits to positive along the minor radius and the outward carbon flux in the negative $E_r$ region.
Since the characteristic $E_r$ profile is obtained regardless of the effect of $\Phi_1$ and without the impact of NBI heating, it is indicated that the high-temperature profiles play a crucial role behind the $E_r$ root-transitioning as discussed in \cite{Velasco2017}.
The results are summarized and discussed in section \ref{sec:disc}.



%% file: Body/DIR_MODEL/Equations.tex
 \subsection{Drift-kinetic equation}

In the global neoclassical model, plasma particles follow the equations of the guiding-center motion in five-dimensional phase space.
As the phase variables, here we choose the guiding-center position $\bm{X}$, the parallel velocity $v_\pll$ and the magnetic moment $\mu=m_a v_\bot^2/(2B)$, where $m_a$ denotes the mass of the species $a$, $v_\bot$ the perpendicular velocity, and $B$ the strength of the magnetic field $\boldsymbol{B}$.
In these coordinates, the equations of motion in static magnetic field are
\begin{eqnarray}
    \label{LJ dX/dt}
\dot{\bm{X}} = &v_\pll\bm{b}
    +\frac{1}{Z_a eB_\pll^\ast}\bm{b}\times\left(
    m_a v_\pll^2 \bm{b}\cdot\nabla\bm{b} + \mu \nabla B+Z_a e\nabla \Phi \right), \\
\label{dU}
\dot{v}_\pll=&-\frac{1}{m_a v_\pll}\dot{\bm{X}}\cdot\left(\mu \nabla B+Z_a e\nabla \Phi\right), \\
\label{LJ dmu/dt}
\dot{\mu}=& 0,
\end{eqnarray}
where $\bm{b}=\bm{B}/B$ is the unit vector along the magnetic field line, $Z_a$ is the charge of the species $a$, $\bm{B}^*=\nabla \times \bm{A}^*$ is the corrected magnetic field with the guiding-center vector potential $\bm{A}^*=\bm{A}+ m_a v_\pll \bm{b}/(Z_a e)$, and $B_\pll^*=\bm{B}^*\cdot \bm{b}$ and the overdot denotes the total time derivative.
Note here that we do not use any approximation based on the order estimation $\Phi_0 \gg \Phi_1$, and therefore $\Phi_1$ is included in all the terms involving $\Phi$ in the equations (\ref{LJ dX/dt}) and (\ref{dU}).
According to the form of the equations (\ref{LJ dX/dt}) and (\ref{dU}), one can see that the relative importance of $\nabla \Phi_1$ term to the magnetic drift and magnetic mirror terms increases for higher $Z_a$.

FORTEC-3D solves the equation for $f_{a1}\equiv f_{a}-f_{a0}$, where $f_{a}$ is the total distribution function of the species $a$ and the analytically known part $f_{a0}$ is given by
\begin{equation}
\label {f0}
f_{a0} = f_{aM}  e^{-Z_a e \Phi_1/T_a},
\end{equation}
with the local Maxwellian
\begin{equation}
    f_{aM}=n_{a0}\left( \frac{m_a}{2\pi T_a} \right)^{3/2} \exp{\left(-\frac{m_a v_\pll^2}{2T_a} - \frac{\mu B}{T_a} \right) },
\end{equation}
where the density $n_{a0}$ and the temperature $T_a$ are flux functions.
As described in \cite{Fujita2020}, $f_{a0}$ does not contribute to the radial fluxes when both the magnetic drift and the $E\times B$-drift due to $\Phi_1$ are considered.

The first order drift-kinetic equation for $f_{a1}$ is then given by
\begin{eqnarray}
\frac{\pd f_{a1}}{\pd t} &+& \dot{\bm{X}}\cdot \nabla f_{a1} + \dot{v}_\pll \frac{\pd f_{a1}}{\pd v_\pll} -C_{TP}(f_{a1})=C_{FP}(f_{a0})
+\frac{Z_a e}{T_a} \frac{\pd \Phi_1}{\pd t}
 \nonumber \\
\label {DKE}
&-&\dot{\bm{X}} \cdot \nabla r
\left [\frac{n_{a0}'}{n_{a0}} +\frac{Z_a e\Phi_0'}{T_a} 
+ \left( \frac{m_a v_\pll^2}{2T_a} + \frac{\mu B}{T_a} -\frac{3}{2} +\frac{Z_a e \Phi_1}{T_a}\right)\frac{T_a'}{T_a}
\right] f_{a0}, 
\end{eqnarray}
where the prime denotes the derivative with respect to the radial coordinate $r$ and $C_{TP}(f_{a1})$ and $C_{FP}(f_{a0})$ are the test particle part and the field particle part of the linearized collision operator, respectively,
\begin{eqnarray}
\label{eq:CTP}
C_{TP}(f_{a1})&=\sum_b C_{TP}^{ab}(f_{a1},f_{b0})
\simeq \sum_b C_{TP}^{ab}(f_{a1},f_{bM}) \\
\label{eq:CFP}
C_{FP}(f_{a0})&=\sum_b C_{FP}^{ab}(f_{a0},f_{b1})
\simeq \sum_b C_{FP}^{ab}(f_{aM},f_{b1}).
\end{eqnarray}
In this work, the conventional Sugama model is used for the collision operator \cite{Satake2020,Sugama2009}.
The recent developments of the collision operator to include higher order terms, which have been neglected but can be important for impurity transport \cite{Calvo2019,Sugama2019}, are not considered.

The equations (\ref{LJ dX/dt})-(\ref{DKE}) are reduced to those for a case without $\Phi_1$ by setting $\Phi_1=0$.
FORTEC-3D employs the two weight $\delta f$ scheme to solve (\ref{DKE}).
See \cite{Fujita2020} and references therein for a derivation of (\ref{DKE}) and the details of the numerical scheme.

%% file: Body/DIR_SETUP/Species-and-EBfields.tex
\subsection{Particle species and electromagnetic fields}

\begin{figure}
    \begin{tabular}{rl}
 \begin{minipage}{0.47\linewidth}
 \centerline{\includegraphics[scale=0.3, angle=-90]
  {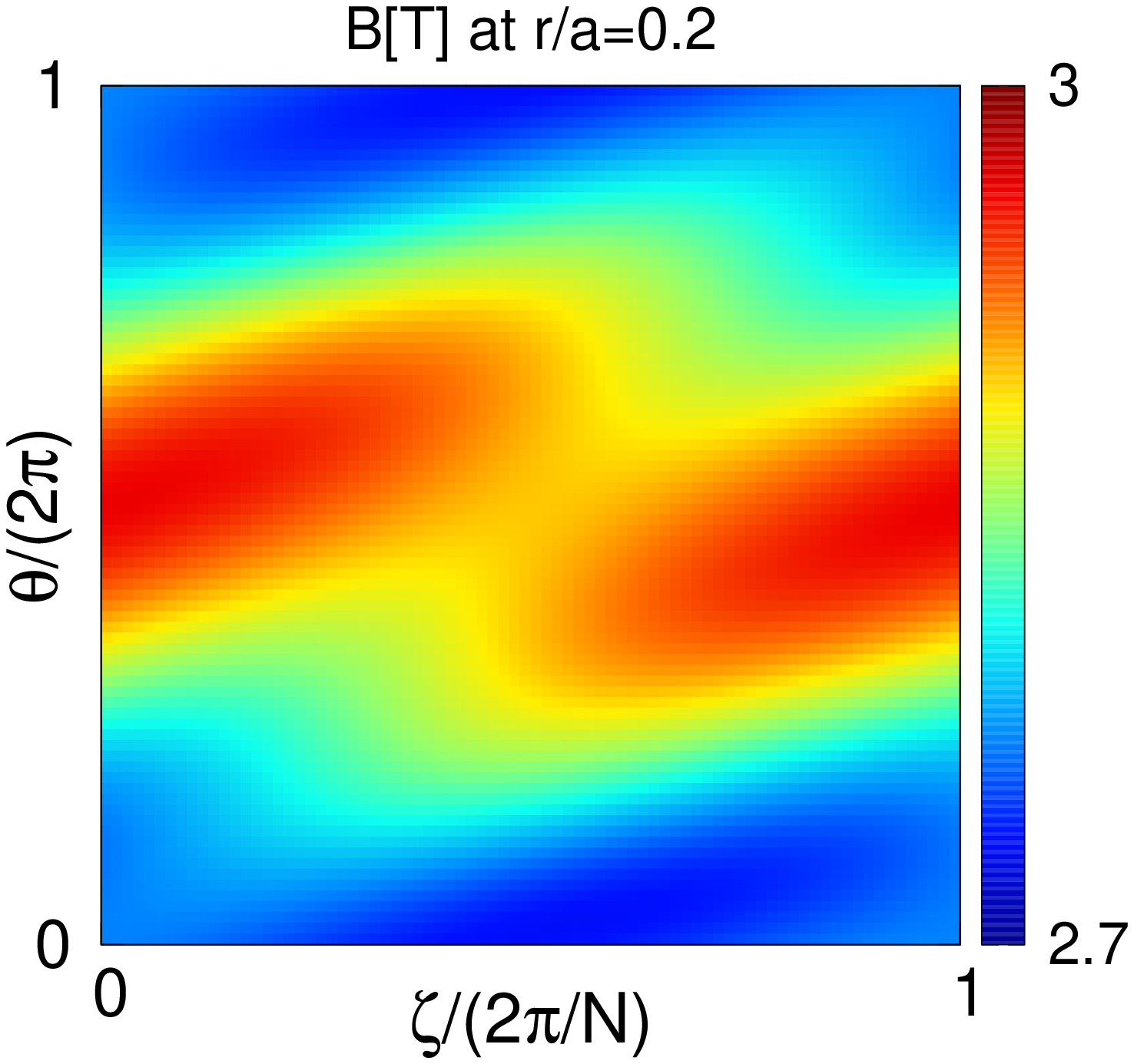}}
 \end{minipage}    
 \begin{minipage}{0.47\linewidth}
 \centerline{\includegraphics[scale=0.3, angle=-90]
  {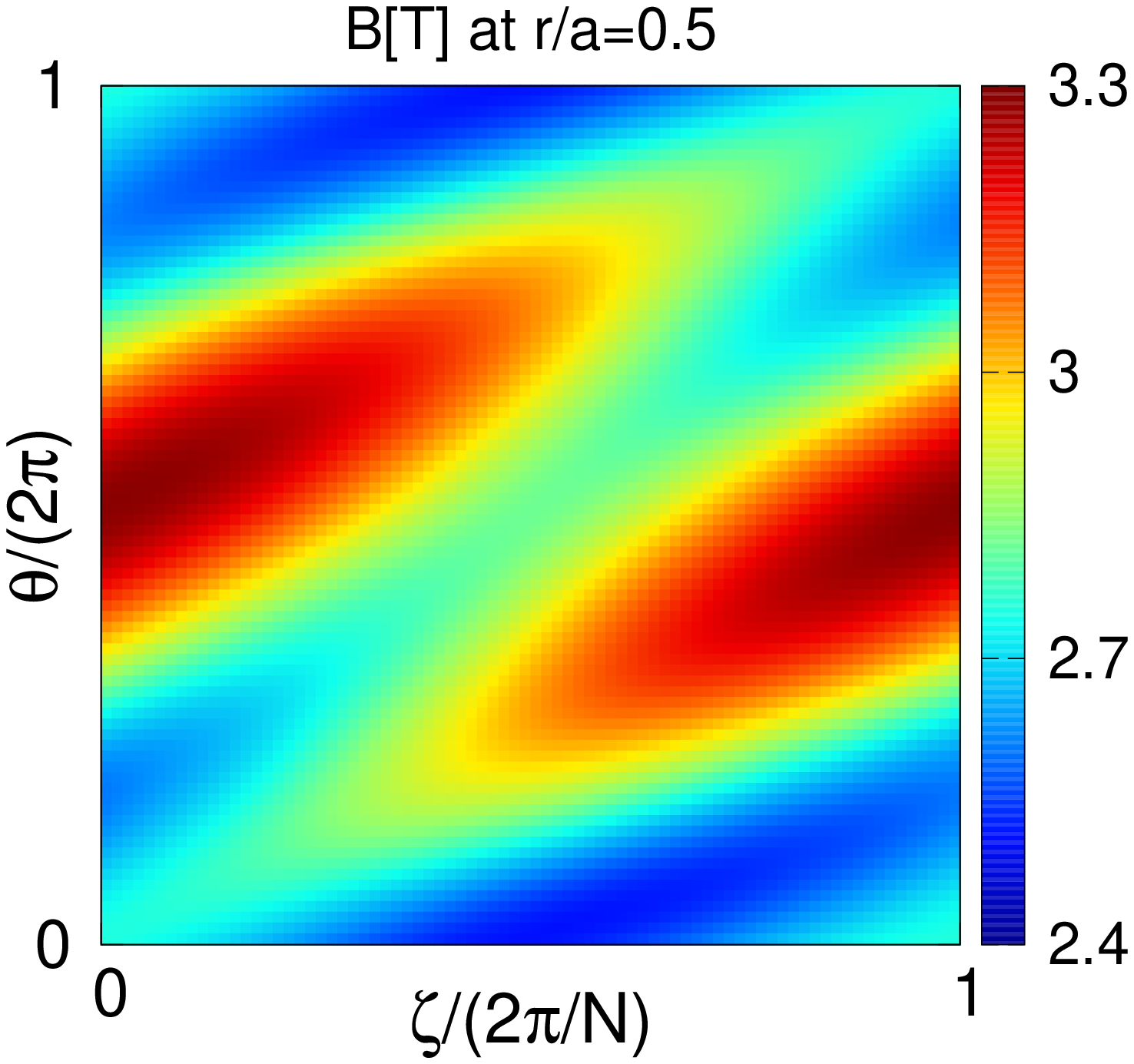}}
 \end{minipage} \\ \\
 \begin{minipage}{0.47\linewidth}
 \centerline{\includegraphics[scale=0.3, angle=-90]
  {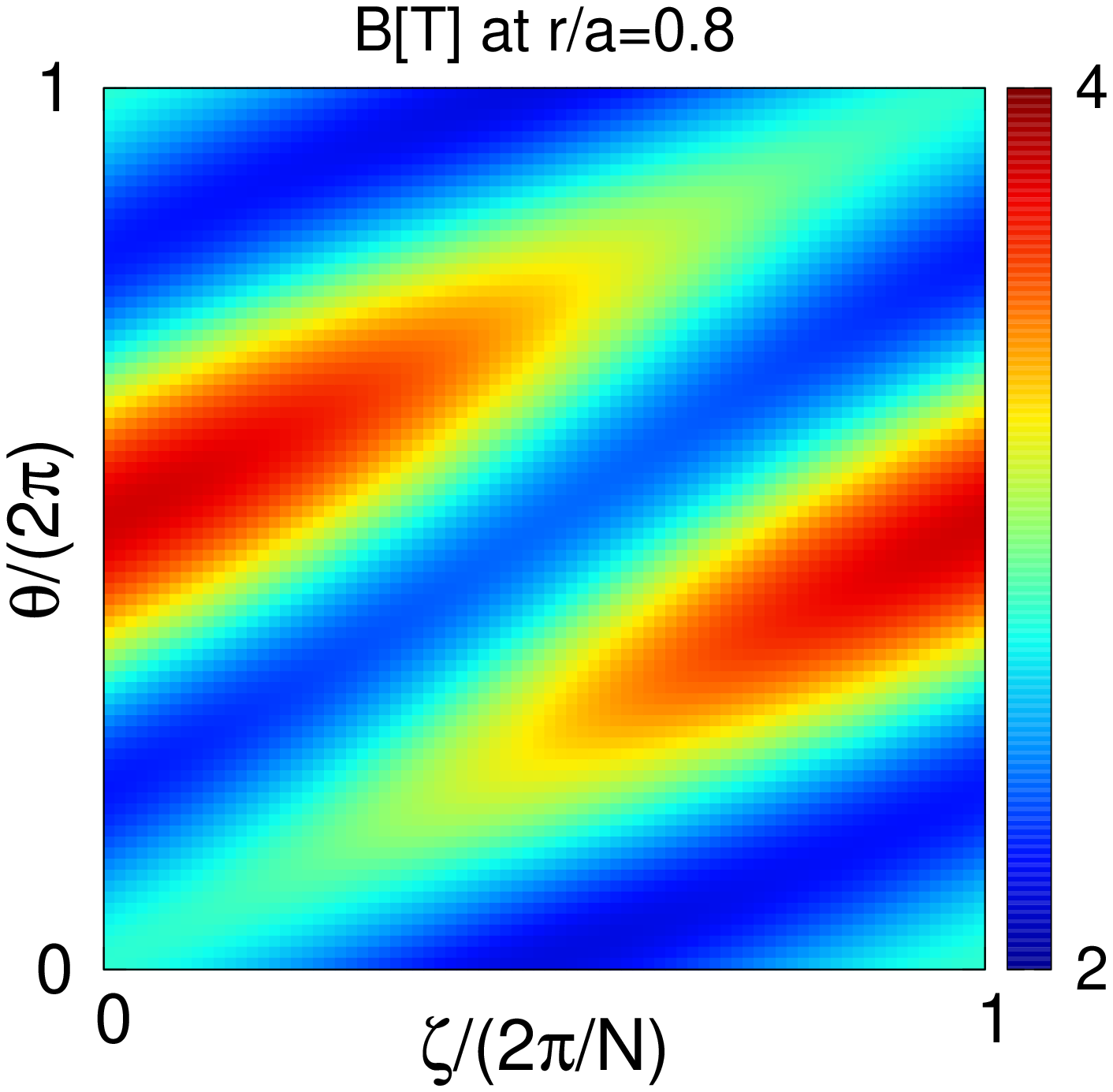}}
 \end{minipage}
 \begin{minipage}{0.47\linewidth}
 \centerline{\includegraphics[scale=0.47]
  {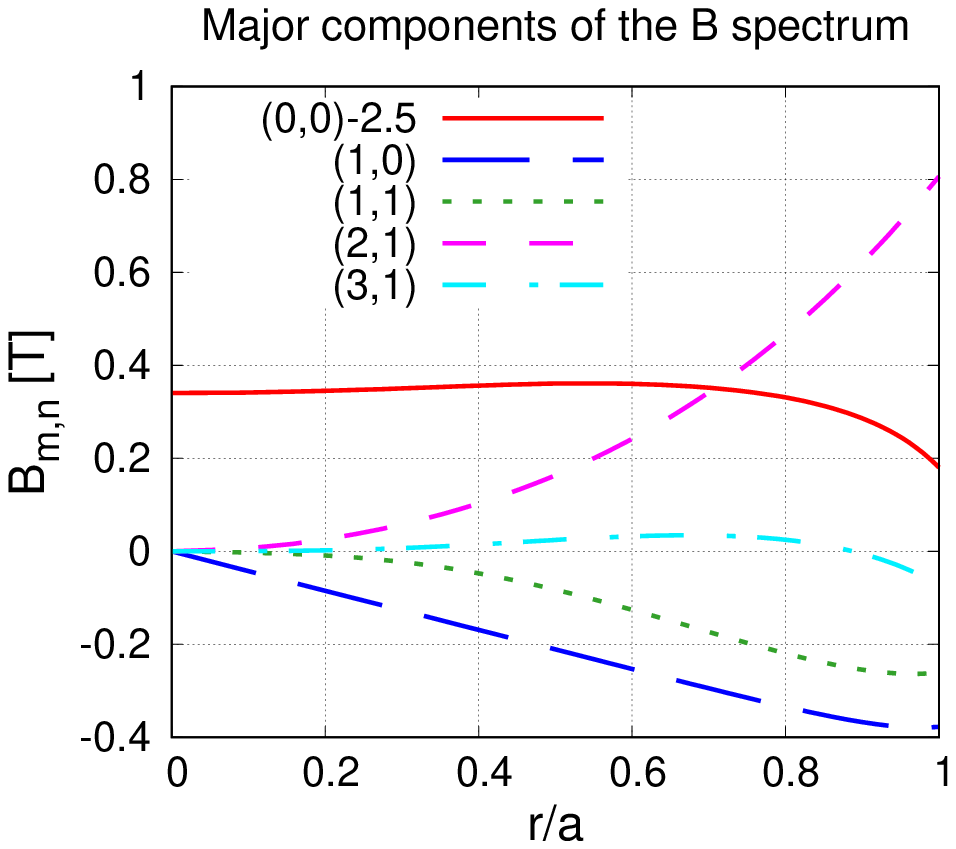}}
 \end{minipage}
    \end{tabular}
      \caption{The magnetic field strength on the flux surface at $r/a=0.2$ (top left), $0.5$ (top right) and $0.8$ (bottom left), and the major Fourier components of the magnetic field (bottom right). 
      The value of the cosine-$(0,0)$ component in the bottom right figure is adjusted by subtracting 2.5 for visualization purposes. 
      $\theta$ and $\zeta$ are the poloidal and the toroidal angles in the Boozer coordinates, respectively, and $N=10$.}
      \label{fig:Bfield}
\end{figure}

In this study, we investigate an LHD plasma with the standard configuration with a major radius of $R_0=3.7$ m and a minor radius of $a=0.62$ m.
The magnetic field strength and its major Fourier components in the Boozer coordinates are represented in Figure \ref{fig:Bfield}.
The plasma contains two different impurity species, helium He$^{2+}$ and carbon C$^{6+}$, as well as the main ion species, hydrogen H$^{1+}$.
Since solving electrons together requires roughly 50 times the numerical cost, the drift-kinetic equation is solved only for the ions by the global code.
Collisions of the ions with the electrons are also neglected. 

When solving the ambipolar condition, electron flux evaluated with a local neoclassical code PENTA \cite{Spong2005} is used.
The $\Phi_1$-effect is not considered in the electron flux.
Also, the contribution of the non-adiabatic variation of the electron density $n_{e1}\equiv \int d^3 v f_{e1}$ is neglected in determining $\Phi_1$.
The first approximation is justified since the impact of $\Phi_1$ on the electron flux is as small as that on the hydrogen flux, which is, as will be shown below, negligible especially in the near-axis region.
The second approximation is by $f_{e1}/f_{i1}\sim \rho_{e}/\rho_{i} \ll 1 $ when $T_i \sim T_e$, where $\rho_a$ denotes the gyro-radius of the species $a$.
However, the validity of the second approximation is questioned when $E_r$ takes large positive values \cite{Garcia2018}.
This point will be briefly discussed in section \ref{sec:disc-Er}.

$\Phi_1$ is determined by the quasi-neutrality condition.
Imposing the quasi-neutrality condition up to the first order and neglecting $n_{e1}$ yields the expression \cite{Garcia2017}
\begin{eqnarray}
    \Phi_1 
    =&\frac{1}{e}\left(\sum_I Z_I^2 \frac{n_{I0}}{T_I}+\frac{n_{e0}}{T_e} \right)^{-1} \sum_I Z_I n_{I1} \nonumber \\
\label{phi1 eq}
    =&\frac{1}{e}\left(\sum_I Z_I^2 \frac{n_{I0}}{T_I}+\frac{n_{e0}}{T_e} \right)^{-1}
    \sum_I Z_I\int d^3v f_{I1},
\end{eqnarray}
where the subscript $I$ refers to ion species.

Time evolution of the ambipolar $E_r$ is determined by
\begin{eqnarray}
\label{eq:evol_Er}
    \epsilon \frac{\pd E_r}{\pd t}
    =
    -e\left( 
    Z_I\sum_I \Gamma_I - \Gamma_e
    \right),
\end{eqnarray}
where $\epsilon$ is the permittivity \cite{Satake2010,Hastings1985} and
\begin{eqnarray}
    \Gamma_a
    \equiv 
    \left\langle 
    \int d^3v \dot{\bm{X}} \cdot \nabla r f_{a1}
    \right\rangle,
\end{eqnarray}
is the radial particle flux of the species $a$ with $\langle ...\rangle$ denoting the flux surface average.
The numerical treatment of the equation  (\ref{eq:evol_Er}) in FORTEC-3D including how to manage with the bifurcation problem is described in \cite{Satake2008}.


In our global simulations, these quantities are evaluated with the initial condition $f_{a1}=E_r=0$. 
The equation for the time evolution of $E_r$ (\ref{eq:evol_Er}) is solved every time step while $\Phi_1$ is updated every certain time steps over which $n_{I1}$ is averaged.
The time averaging length is about $1/40$ times the collision time.


%% file: Body/DIR_SETUP/nTprof.tex
\subsection{Plasma profiles}

\begin{figure}
 \centerline{\includegraphics[scale=0.6, angle=-90]
  {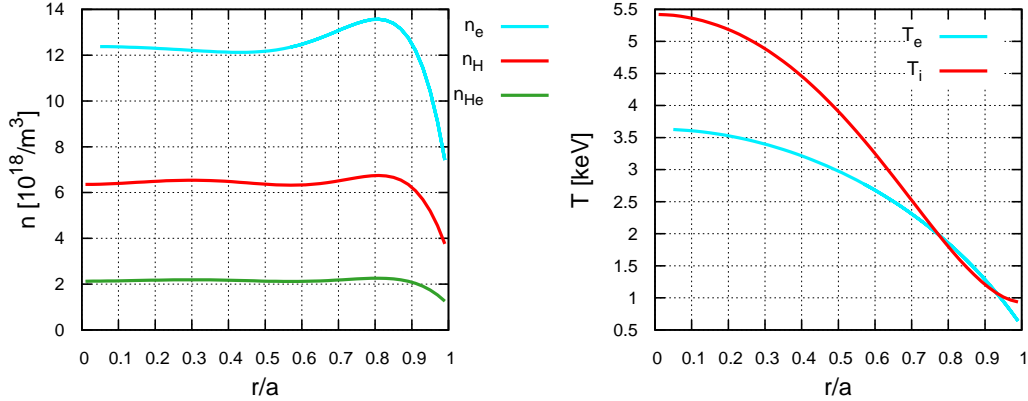}}
      \caption{The radial profiles of the densities (left) and temperatures (right) except for the carbon density profile.}
      \label{fig:nT_prof_I}
\end{figure}

We investigate three different cases, each corresponding to a different carbon density profile.
The $n$-$T$ profiles, including the carbon density profile, for case A are the same as those used in  previous studies on an impurity hole plasma observed in LHD \cite{Mollen2018,Nunami2020,Fujita2020}.
All of the ion species are assumed to be in thermal equilibrium with each other so that they have the same temperature $T_I=T_i$.
That which characterizes an impurity hole plasma and distinguishes it from other ion-root plasmas is its high ion temperature in the core induced by NBI. 
The radial profiles of the temperatures and densities except for the carbon density profile are plotted in Figure \ref{fig:nT_prof_I}.

\begin{figure}
 \centerline{\includegraphics[scale=0.6, angle=-90]
  {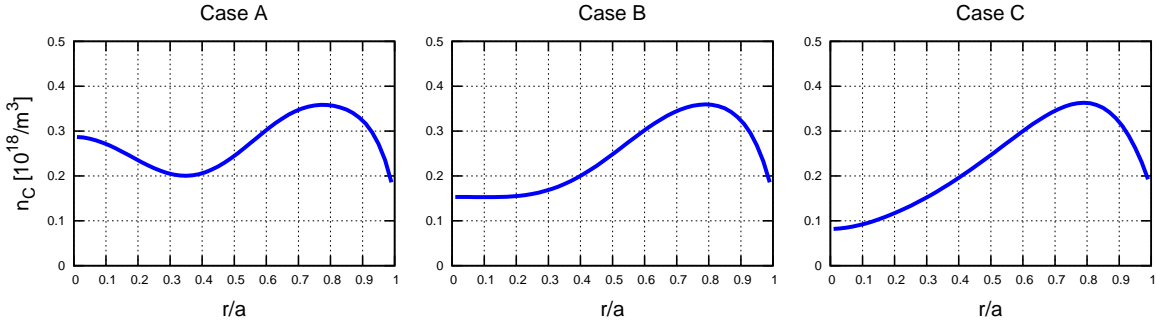}}
      \caption{The radial carbon density profiles for case A (left), B (center) and C (right), respectively.}
      \label{fig:nC}
\end{figure}

\begin{figure}
 \centerline{\includegraphics[scale=0.6, angle=-90]
  {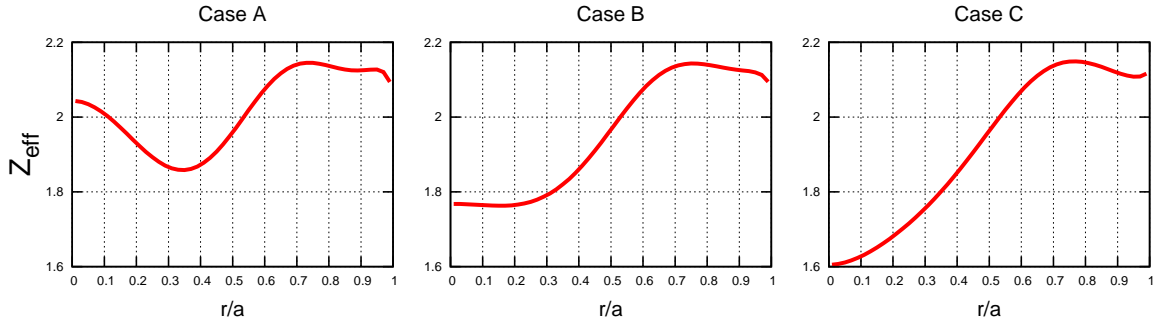}}
      \caption{The radial profiles of $Z_{\rm eff}$ for case A (left), B (center) and C (right), respectively.}
      \label{fig:Zeff}
\end{figure}

The carbon density profiles for each case are plotted in Figure \ref{fig:nC} and the corresponding profiles of the effective charge $Z_{\rm eff}\equiv \sum_I n_I Z_I^2/n_e$ are shown in Figure \ref{fig:Zeff}.

For case A, the hollow structure of the carbon density profile is formed at an off-axis region ($0.3<r/a<0.7$). 
The density gradient near the axis thus can contribute to driving the carbon impurities outwardly. 
Some measurement data, however, indicate that impurity holes are formed around the magnetic axis \cite{Ida2009,Yoshinuma2009, Ido2010}. 
For the rest of the cases, the carbon density profile is thus modified to be hollow around the magnetic axis. 
For case B, the carbon density gradient near the axis is flattened.
On the other hand, the gradient near the axis is steepened for case C.
Thus, while the density gradient $dn_c/dr$ does not work as a driving force for case B, it works as an inward force for the carbon flux for case C.
The carbon impurities in this plasma are in the plateau regime for the entire radius (see figure 4 in \cite{Fujita2020}). 

In order to preserve the charge-neutrality, the electron density is modified accordingly to the modification in the carbon density profiles as well.
However, the differences in the electron density profiles are not explicitly shown because this level of difference can be neglected.
The ion and electron temperatures are fixed since it is observed that the changes in the temperatures during phase when the carbon density profile becomes hollower are also small \cite{Ida2009,Mikkelsen2014,Yoshinuma2010}.

In order to study the effect of $\Phi_1$, we simulate the cases with and without $\Phi_1$ for each case.

%% file: Body/DIR_RESULTS/ResultA.tex
\subsection{Case A}\label{sec:resultA}

\begin{figure}
 \centerline{\includegraphics[scale=0.3, angle=-90]
  {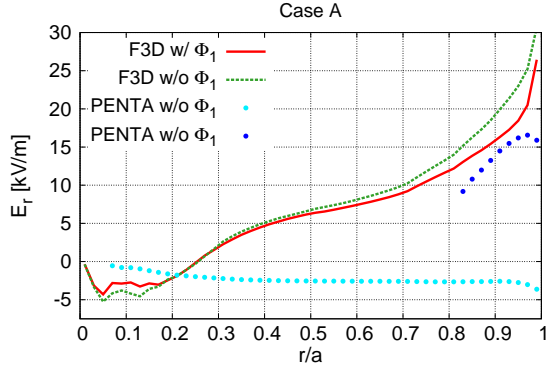}}
      \caption{The ambipolar radial electric field profiles for case A.
      The red and the green lines correspond to the results of FORTEC-3D with and without $\Phi_1$, and the cyan and the blue points to PENTA results, respectively.}
      \label{fig:Er_A} 
\end{figure}

\begin{figure}
 \centerline{\includegraphics[scale=0.3, angle=-90]
  {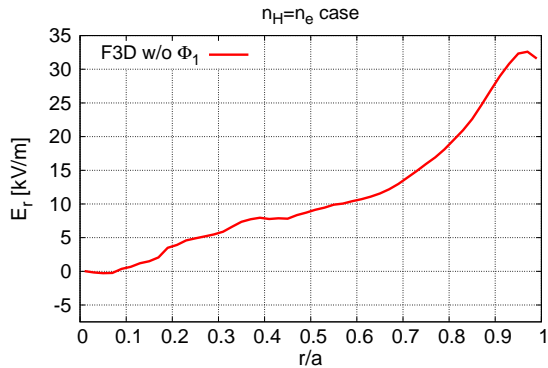}}
      \caption{The ambipolar radial electric field profiles obtained by global simulation assuming $n_H=n_e$.}
      \label{fig:Er_pure}
\end{figure}

For case A, the profile of the ambipolar $E_r$ obtained by the global simulation turns out to be quite different from those obtained by local simulations (Figure \ref{fig:Er_A}).
The cyan points represent the solution for the case without $\Phi_1$ obtained by using a local code PENTA.
The sign of the local solution is negative  for almost the entire radius.
The local calculation also finds an electron root, represented with blue points, but only partially at $r/a>0.8$.
On the other hand, the sign of the global solutions, which are represented with the red and green lines, is negative near the magnetic axis but transits to positive around $r/a=0.25$.
The red and green lines correspond to the cases with and without $\Phi_1$, respectively.
The solution for a pure plasma, in which the electron density $n_e$ is equal to the hydrogen density $n_H$, is also shown in Figure \ref{fig:Er_pure}.
By comparing Figure \ref{fig:Er_A} and Figure \ref{fig:Er_pure}, we find the $E_r$ value is lowered maintaining its profile shape when the impurity contributions are taken into account.
The resulting emergence of the ion-root near the axis is a distinctive feature of the simulation result and indicates that the impurity contributions are not negligible in the ambipolar condition.

Compared with the difference between the global solutions, the differences between the global and the local solutions are substantial.
However, it is known that the condition of an impurity hole plasma is, even if it were in ion-root, very close to the condition for $E_r$ to be positive \cite{Velasco2017}.
The existence of the positive local solution near the edge suggests that the cause of the  difference is more subtle than it looks at a glance.
Also, and more crucially, the sign-changing feature is shared with the experimentally observed profile \cite{Ido2010} as already mentioned in section \ref{sec:Intro}. 
Note also that positive $E_r$ in the core region of $T_i >T_e$ discharges have been observed in LHD high-$T_i$ and low-$n_e$ shots \cite{Nagaoka2015}, where even a local neoclassical code predicts the core electron-root.
The difference in shapes of the positive $E_r$ profiles evaluated with FORTEC-3D and PENTA, respectively, at the edge region ($r/a>0.95$) can be caused by several differences in the models, such as in the orbits and the orbit loss treatments, and in the collision operators.
Nevertheless, both codes are designed so that the treatments on the edge boundary do not affect the physics in the inner region. 

One may be concerned with the poloidal Mach number $M_p\simeq  E_rR_0/(v_{th} B_0 r \iota)$ for carbon, where $B_0$ is the magnetic field strength at the magnetic axis, $v_{th}$ is the thermal velocity and $\iota$ is the rotational transform, respectively, since nonlinear dependence of the neoclassical poloidal viscosity (and therefore of the radial flux) appears at $|M_p|\simeq 1$, which is caused by the resonance of parallel motion and $E\times B$ rotation \cite{Shaing1993,Dahi1998}. 
For the ambipolar $E_r$ profile in Figure \ref{fig:Er_A}, $|M_p|$ for carbon is $0.2
\sim 0.4$ except for very close to the magnetic axis $r/a <0.05$, where $|M_p|>1$. 
Therefore, the resonance effect is basically irrelevant to the carbon impurity neoclassical flux in the present study.

\begin{figure}
 \centerline{\includegraphics[scale=0.7, angle=-90]
  {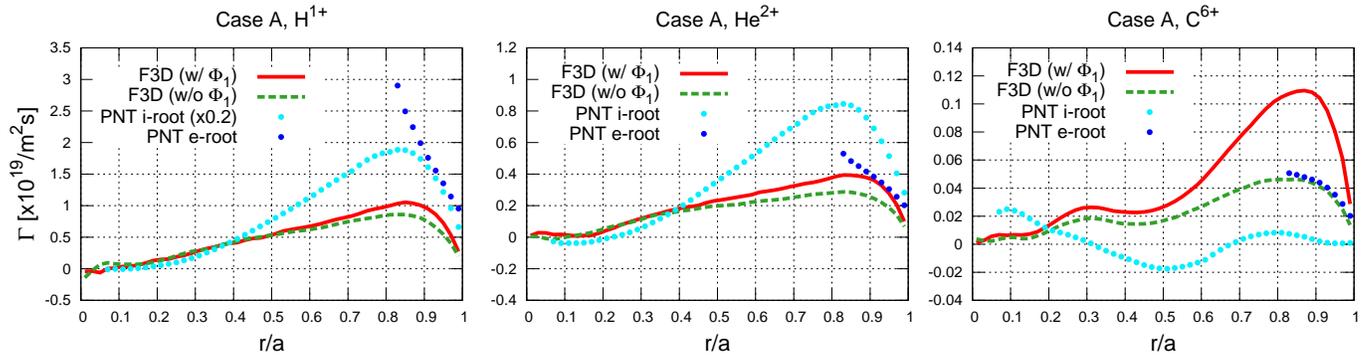}}
      \caption{The radial fluxes of H$^{1+}$ (left), He$^{2+}$ (center) and C$^{6+}$ (right) for case A.
      The red and the green lines correspond to the global calculation results with and without $\Phi_1$, and the cyan points and the blue points to the PENTA results of the ion-root and the electron-root cases, respectively.
      Note that the hydrogen flux of the PENTA calculation for the ion-root case is plotted being multiplied by $0.2$.}
      \label{fig:flux_A}
\end{figure}

In Figure \ref{fig:flux_A}, the radial particle fluxes of the ion species are plotted.
As in Figure \ref{fig:Er_A}, the red and green lines correspond to the global calculation results with and without $\Phi_1$, respectively.
For reference, the PENTA results with the ion-root and the electron-root of the ambipolar $E_r$ are also plotted with cyan points and blue points, respectively. 
Note that the value of the hydrogen flux for the local ion-root case is multiplied by $0.2$ for visualization purposes. 
That is, the original value of the hydrogen flux is five times larger than represented.
The reason that the discrepancy in the flux values between the global and local models becomes larger for lighter species is that the impact of the magnetic drift becomes largest for low collisionality regimes \cite{Matsuoka2015,Fujita2020,VelascoKNOSOS}.
For this case, briefly, the absence of the magnetic drift leads to the overestimation of the $1/\nu$-type trapped particle diffusion where $E_r$ is close to zero since the particles are stuck in the local helical ripple wells without drift motion.
Inclusion of the precession motion of the trapped particles by the magnetic drift significantly moderates the excessive diffusion. 
The difference in the ambipolar $E_r$ profiles between global and local simulations reflects whether or not the magnetic drift is considered.

As can be seen, regardless of the effect of $\Phi_1$, all of the ion fluxes are outwardly directed for almost the entire radius.
This is the case even for the carbon flux where the sign of $E_r$ is negative ($r/a<0.25$) and where the carbon density profile is hollow (from $r/a\sim 0.3$ to $r/a\sim 0.7$).
The effect of $\Phi_1$ tends to drive the flux more outwardly and its impact is largest for carbon.
Therefore, considering the impurity transport in the ambipolar condition increases the portion of the escaping positive charge.
The $E_r$ level is consequently lowered as the shift from Figure \ref{fig:Er_pure} to Figure \ref{fig:Er_A}.


Note that the carbon flux for the local ion-root case is also outwardly directed near the magnetic axis.
This is seen in a previous study as well \cite{Mollen2018}.
Remember that for this case, however, the carbon density profile is peaked around the axis.
In order for the flux to be consistent with the carbon density profile, the flux should be positive where the hole structure is formed, but this is not the case. 
To the contrary, the global results are consistent with the density profile as described above.

To be exact, the profile of the turbulent flux, not only the neoclassical flux, is needed to discuss the consistency.
Still, the gyrokinetic study has shown that the values of the global neoclassical fluxes of the ion species are close to those to balance with the turbulent fluxes whereas the local neoclassical fluxes are out of balance. 
The existing data, though still insufficient, are thus in favor of our result on this point as well.
We will return to this point in section \ref{sec:disc:balance}.

\begin{figure}
 \centerline{\includegraphics[scale=0.65, angle=-90]
  {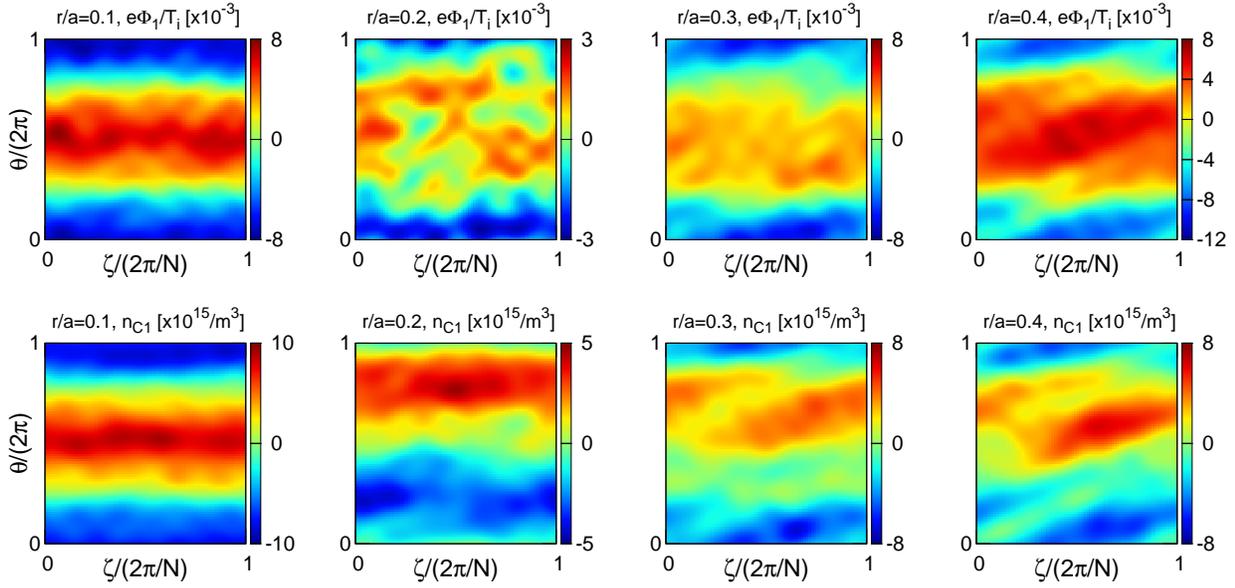}}
      \caption{The spatial structures of $\Phi_1$ (the upper figures) and $n_{C1}$ (the lower figures) for case A.
      From left to right, each column corresponds to the flux surface at $r/a=0.1, 0.2, 0.3$ and $0.4$, respectively.}
      \label{fig:Phi1_A}
\end{figure}

Figure \ref{fig:Phi1_A} shows the 2-dimensional spatial structures of $\Phi_1$ and the carbon density variation $n_{C1}$ on the flux surfaces at $r/a=0.1, 0.2, 0.3$ and $0.4$, respectively.
$\theta$ and $\zeta$ are the poloidal and the toroidal angles of the Boozer coordinates, respectively, and $N$ is the toroidal symmetry number ($N=10$ for LHD).
The poloidal angle $\theta$ is chosen to be zero on the outboard side.
The phase of $\Phi_1$, which is mainly determined by the phase of the hydrogen density variation $n_{H1}$, is stellarator symmetric while the phase of carbon density appears to be stellarator anti-symmetric.
These results are consistent with analytical estimation of the collisionality dependence of the size and the phase of density variation, and therefore of $\Phi_1$ \cite{Calvo2017,Alonso2017}.
The structural difference in the density variations can roughly be argued by a criterion known for classical stellarators as well.
When the poloidal precession frequency $\Omega_\theta\simeq \Omega_E\simeq  E_r/(rB_{00})$ exceeds the effective collision frequency $\nu_{\rm eff}$, the $\cos\theta$ component becomes the leading mode in the spatial distribution.
On the other hand, as $\nu_{\rm eff}/\Omega_\theta$ increases, the $\sin\theta$ component grows relatively larger \cite{Garcia2017,Beidler1995}.
For this case, $\nu_{\rm eff}/\Omega_\theta\sim O(1)$ for carbon while $\nu_{\rm eff}/\Omega_\theta \sim O(0.1)$ for hydrogen and helium.

\begin{figure}
 \centerline{\includegraphics[scale=0.6, angle=-90]
  {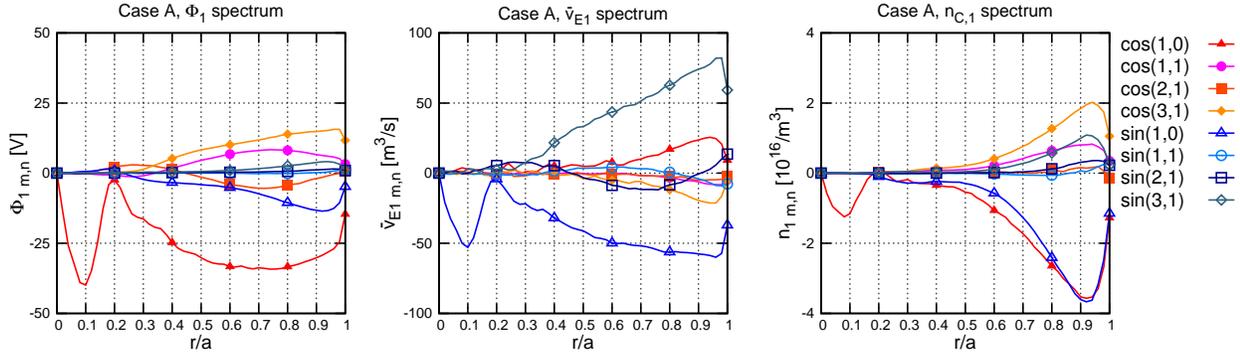}}
      \caption{The radial profiles of the leading modes in the Fourier spectra of $\Phi_1$ (left), $\bar{v}_{E1}$ (center) and $n_{C1}$ for case A, respectively.}
      \label{fig:spectra_A}
\end{figure}

Among the presented figures, the amplitude of both $\Phi_1$ and $n_{C1}$ become minimum at $r/a=0.2$, which is a position close to the transition point of $E_r$.
Also, the structure of $\Phi_1$ on the surface at $r/a=0,2$ is rather disordered compared with those on the other surfaces.
Figure \ref{fig:spectra_A} shows this in terms of the Fourier spectrum.
The left figure is the radial profile of the $\Phi_1$ spectrum.
As can be seen, the leading mode, $\cos{(1,0)}=\cos\theta$, is shrunk around $r/a=0.2$ and no single specific mode becomes dominant there.
The center figure shows the spectrum of
\begin{eqnarray}
    \bar{v}_{E1} \equiv \sqrt{g} \bm{v}_{E1}\cdot \nabla r,
\end{eqnarray}
where $\sqrt{g}$ is the Jacobian and $\bm{v}_{E1}$ is the $E\times B$-drift generated by $\Phi_1$.
Not $\bm{v}_{E1}$ itself but $\bar{v}_{E1}$ is plotted since the radial particle flux driven by $\bm{v}_{E1}$ can be calculated by the sum of the products of the Fourier coefficients of $\bar{v}_{E1}$ and $n_{a1}$ \cite{Fujita2020}.
Since $\bm{v}_{E1}\cdot \nabla r$ is given by a derivative of $\Phi_1$, the $\sin{(1,0)}(=\sin\theta)$ mode becomes one of the dominant modes in the $\bar{v}_{E1}$ spectrum.
As shown in the right figure, this mode becomes dominant in the $n_{C1}$ spectrum at $r/a>0.2$ as well.
The coupling of this mode with the same sign between $\bar{v}_{E1}$ and $n_{C1}$ results in the outward enhancement of the carbon flux at outer radii (see Figure \ref{fig:flux_A}).

%% file: Body/DIR_RESULTS/ResultB.tex
\subsection{Case B}

The realization of the outward carbon flux near the axis for case A is not so surprising since the carbon density gradient at the region is negative, which works as an outward driving force, and the $E_r$ takes positive value where the density gradient is no longer negative.
For case B, the outward driving force is removed by flattening the carbon density profile near the axis.
Let us see how the behaviors of the carbon impurities respond to this modification. 

\begin{figure}
 \centerline{\includegraphics[scale=0.3, angle=-90]
  {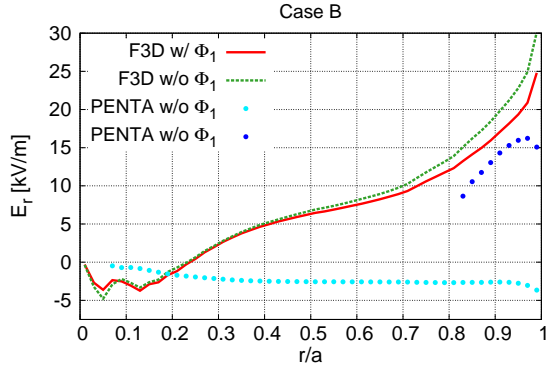}}
      \caption{The ambipolar radial electric field profiles for case B.
      The red and the green lines correspond to the results of FORTEC-3D with and without $\Phi_1$, and the cyan and the blue points to PENTA results, respectively.}
      \label{fig:Er_B}
\end{figure}

\begin{figure}
 \centerline{\includegraphics[scale=0.7, angle=-90]
  {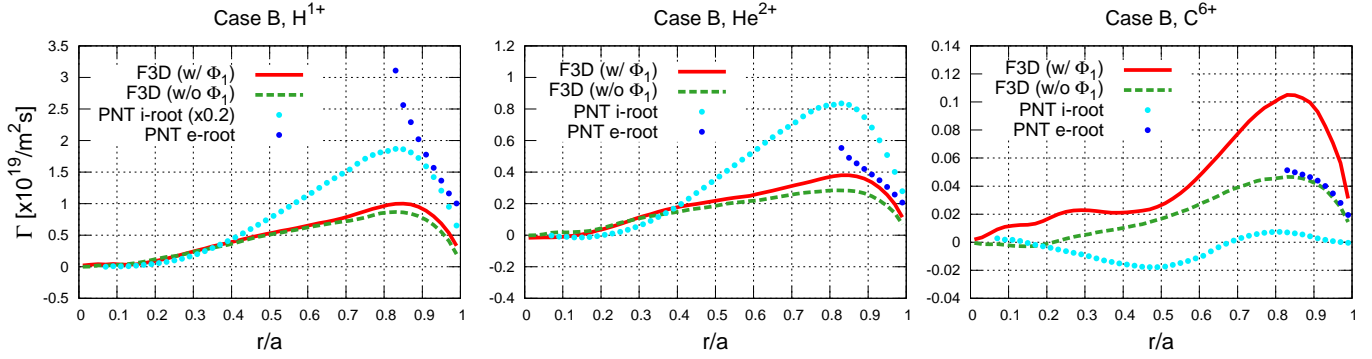}}
      \caption{The radial fluxes of H$^{1+}$ (left), He$^{2+}$ (center) and C$^{6+}$ (right) for case B.
      The red and green lines correspond to the global calculation results with and without $\Phi_1$, and the cyan points and the blue points to the PENTA results of the ion-root and the electron-root cases, respectively.
      Note that the hydrogen flux of the PENTA calculation for the ion-root case is plotted being multiplied by $0.2$.}
      \label{fig:flux_B}
\end{figure}

The ambipolar $E_r$ profile for case B is close to that for case A (Figure \ref{fig:Er_B}).
The $E_r$ thus contributes to driving the ion fluxes inwardly near the axis. 
Further, unlike case A, the carbon density gradient near the axis is flat for this case.
As a consequence, the carbon flux without $\Phi_1$ is slightly negative around $0.1< r/a <0.2$ though its absolute value is close to zero as shown in Figure \ref{fig:flux_B}.
However, when $\Phi_1$ is taken into account, the flux is driven outwardly for the entire radius.
The flux profiles of the other ion species are very similar to those for case A.
As shown in the analysis of \cite{Velasco2018,Velasco2017}, the outward contribution of the ion temperature gradient to the carbon flux can be comparable with the inward contribution of the negative but small $E_r$ in high-$T_i$ plasmas.
The almost zero carbon flux near the axis for the case without $\Phi_1$ manifests this feature.

\begin{figure}
 \centerline{\includegraphics[scale=0.65, angle=-90]
  {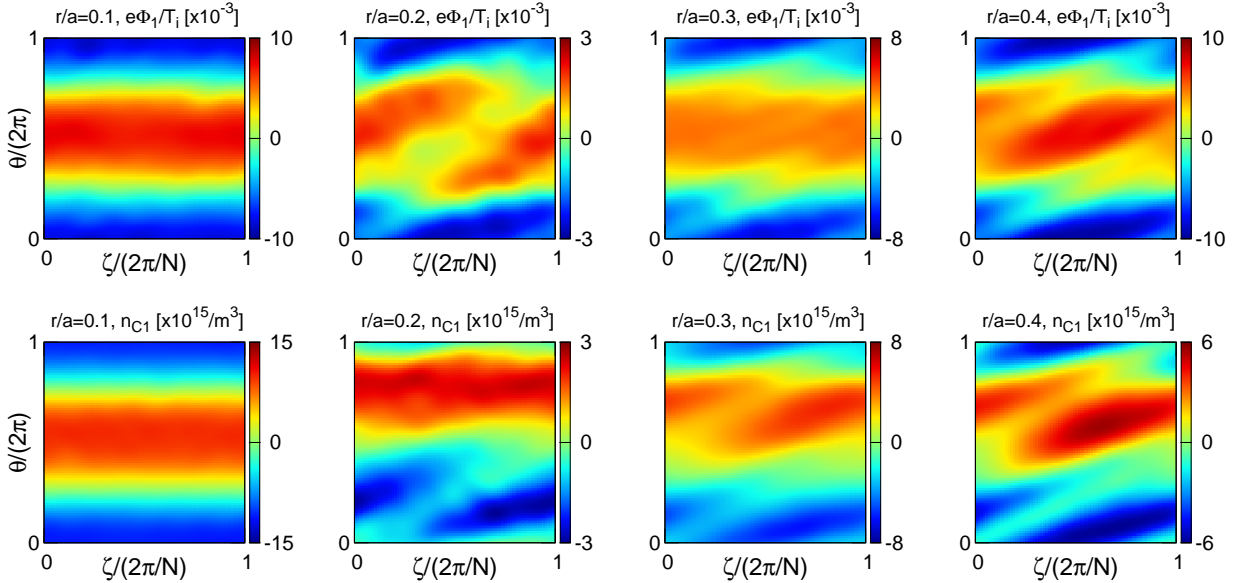}}
      \caption{The spatial structures of $\Phi_1$ (the upper figures) and $n_{C1}$ (the lower figures) for case B.
      From left to right, each column corresponds to the flux surface at $r/a=0.1, 0.2, 0.3$ and $0.4$, respectively.}
      \label{fig:Phi1_B}
\end{figure}

\begin{figure}
 \centerline{\includegraphics[scale=0.6, angle=-90]
  {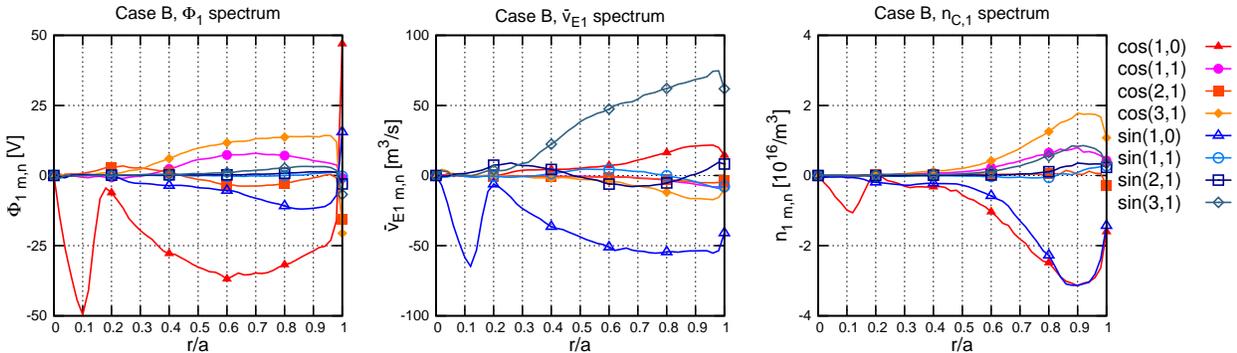}}
      \caption{The radial profiles of the leading modes in the Fourier spectra of $\Phi_1$ (left), $\bar{v}_{E1}$ (center) and $n_{C1}$ for case B, respectively.}
      \label{fig:spectra_B}
\end{figure}

The reason for the outward enhancement of the carbon flux can be seen in the structures of $\Phi_1$ and $n_{C1}$ (Figure \ref{fig:Phi1_B}) and their Fourier spectra (Figure \ref{fig:spectra_B}).
While the $\Phi_1$ spectrum, and therefore the $\bar{v}_{E1}$ spectrum, for this case are similar to those for case A, a subtle difference in the $n_{C1}$ spectrum can be seen.
The $\sin(1,0)$ mode in the $n_{C1}$ spectrum starts growing around $r/a=0.1$ and couples with the same mode in the $\bar{v}_{E1}$ spectrum.
In fact, although it is too small to see in the figure, the $\sin(1,0)$ mode takes finite negative values near the axis as well.
The $\sin(1,0)$ mode with the negative sign is also a leading mode in the radial magnetic drift $\bm{v}_m\cdot \nabla r$.
These couplings result in the outward carbon flux for the entire radius.

%% file: Body/DIR_RESULTS/ResultC.tex
\subsection{Case C}

For case B, we saw that the carbon flux near the axis can be outwardly directed even where $dn_{C}/dr\sim 0$ and $E_r<0$. 
For case C, the carbon density gradient is further steepened and rendered to have positive values.

\begin{figure}
 \centerline{\includegraphics[scale=0.3, angle=-90]
  {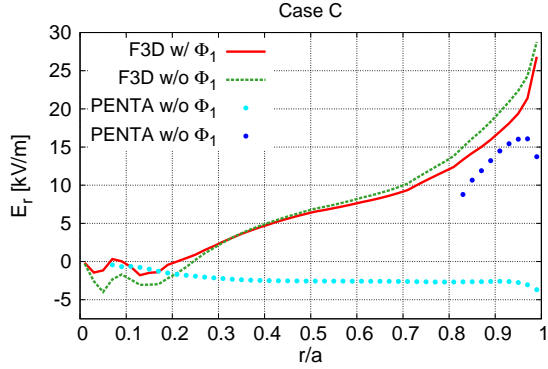}}
      \caption{The ambipolar radial electric field profiles for case C.
      The red and the green lines correspond to the results of FORTEC-3D with and without $\Phi_1$, and the cyan and the blue points to PENTA results, respectively.}
      \label{fig:Er_C}
\end{figure}

The ambipolar $E_r$ profile for case C without $\Phi_1$ is qualitatively analogous to those for the first two cases.
However, when $\Phi_1$ is included, the value of $E_r$ is shifted to be close to zero near the axis (Figure \ref{fig:Er_C}).

\begin{figure}
 \centerline{\includegraphics[scale=0.7, angle=-90]
  {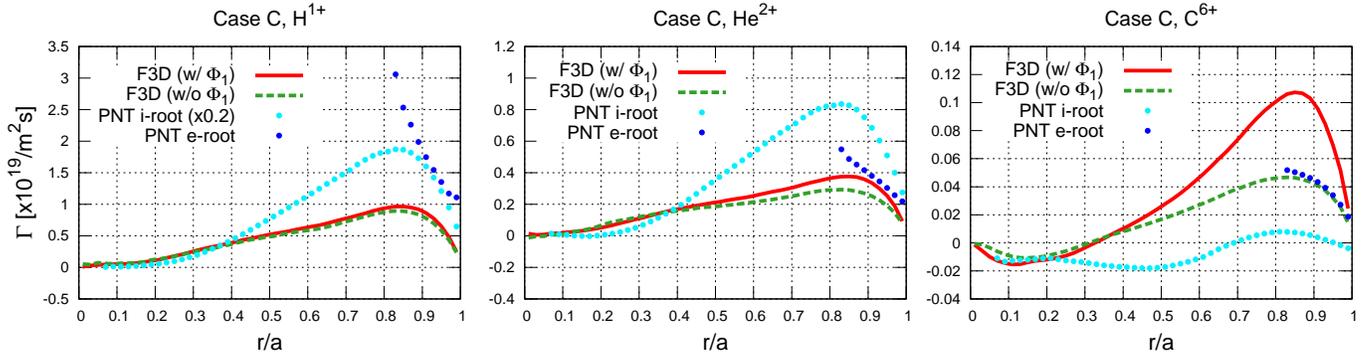}}
      \caption{The radial fluxes of H$^{1+}$ (left), He$^{2+}$ (center) and C$^{6+}$ (right) for case C.
      The red and green lines correspond to the global calculation results with and without $\Phi_1$, and the cyan points and the blue points to the PENTA results of the ion-root and the electron-root cases, respectively.
      Note that the hydrogen flux of the PENTA calculation for the ion-root case is plotted being multiplied by $0.2$.}
      \label{fig:flux_C}
\end{figure}

In Figure \ref{fig:flux_C}, the radial profiles of the ion particle fluxes are represented.
The hydrogen and helium fluxes are also similar to those for the previous two cases and are insensitive to the change in the $E_r$ profile near the axis due to $\Phi_1$.
In contrast, the carbon flux near the axis is inwardly directed whether or not $\Phi_1$ is included.
This indicates that the steep gradient of the carbon density becomes a dominant driving force for the carbon flux.
Also, unlike the previous two cases, $\Phi_1$ contributes to driving the carbon flux more inwardly. 

\begin{figure}
 \centerline{\includegraphics[scale=0.65, angle=-90]
  {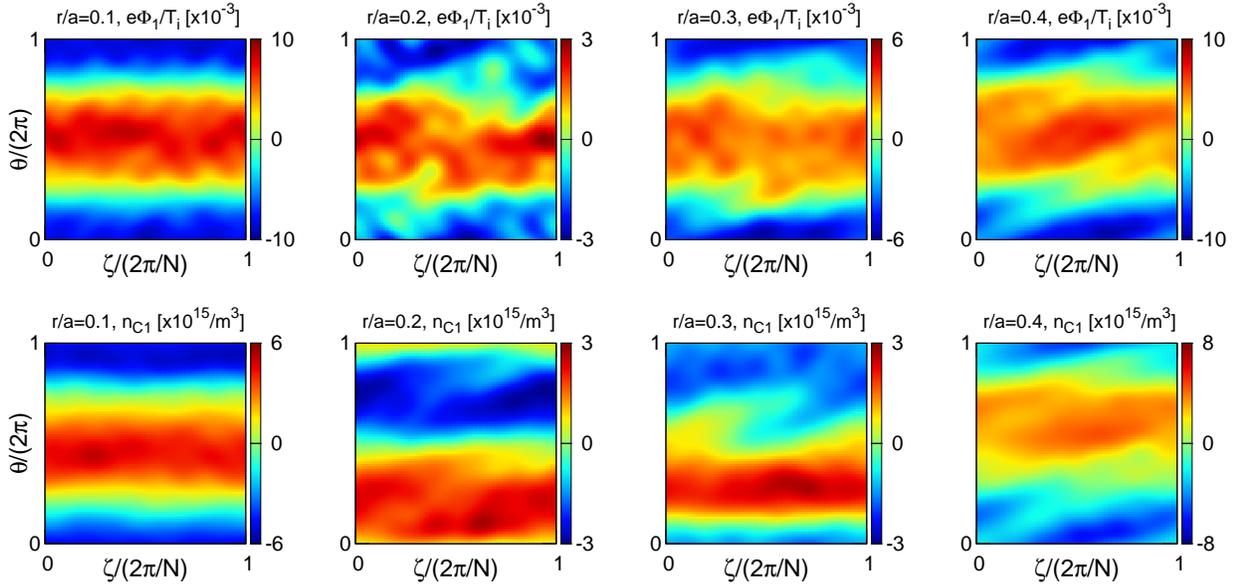}}
      \caption{The spatial structures of $\Phi_1$ (the upper figures) and $n_{C1}$ (the lower figures) for case C.
      From left to right, each column corresponds to the flux surface at $r/a=0.1, 0.2, 0.3$ and $0.4$, respectively.}
      \label{fig:Phi1_C}
\end{figure}

\begin{figure}
 \centerline{\includegraphics[scale=0.6, angle=-90]
  {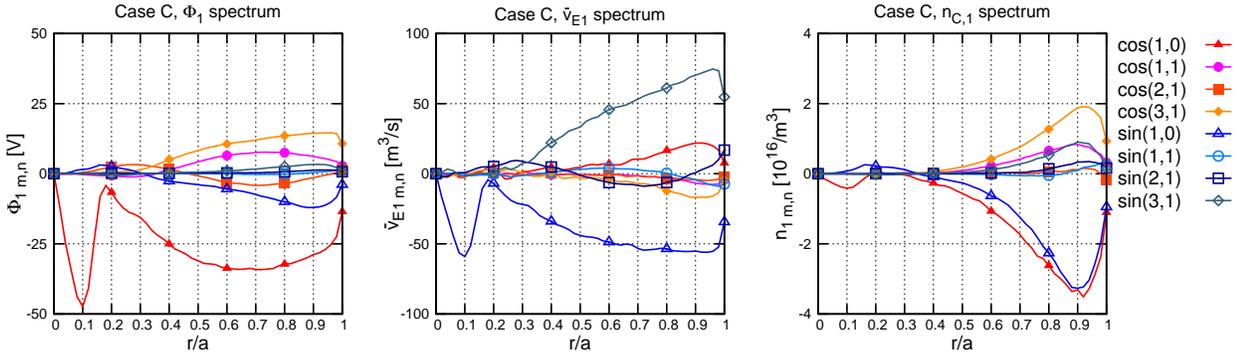}}
      \caption{The radial profiles of the leading modes in the Fourier spectra of $\Phi_1$ (left), $\bar{v}_{E1}$ (center) and $n_{C1}$ for case C, respectively.}
      \label{fig:spectra_C}
\end{figure}

This inward enhancement is also caused by the coupling of the $\sin(1,0)$ mode. 
The spatial structures of $\Phi_1$ and $n_{C1}$ are represented by Figure \ref{fig:Phi1_C}.
By comparing the $n_{C1}$ structure at $r/a=0.2$ with that for case B (Figure \ref{fig:Phi1_B}), it can be seen that the distribution of the carbon impurities is inverted in $\theta$-direction. 
This reflects the fact that the sign of the $\sin(1,0)$ mode at $r/a<0.3$ for this case is opposite from that for case B (Figure \ref{fig:spectra_C}).
This inversion results in the opposite-sign coupling between $\bar{v}_{E1}$ and $n_{C1}$ and leads to the inward enhancement of the carbon flux near the axis.

%% file: Body/Discussion.tex
 \subsection{Global effects and ambipolar radial electric field profile}\label{sec:disc-Er}

By global simulation, we found the solutions of the ambipolar condition which change the sign from negative to positive along the minor radius.
With such $E_r$, the impurity carbon flux can be outwardly directed even where $E_r<0$ and the carbon density profile is hollow.
The root-transitioning has been experimentally observed in an impurity hole plasma and our result is thus qualitatively consistent with the experiment.
However, several tasks and challenges still remain to reveal the mechanism behind the impurity hole formation.

By global neoclassical simulation, a radially continuous profile of ambipolar $E_r$ and of the corresponding neoclassical fluxes can be obtained even when the $E_r$ changes its sign.
By local simulation, on the other hand, the connection between an ion root and an electron root cannot be determined by itself.
To determine how to connect the two different roots by local simulation, the finite-orbit-width (FOW) effect or the effects of anomalous transport on the radial fluxes need to be modelled and incorporated into the simulation model \cite{Hastings1985dif,Itoh2001}.
The resulting $E_r$ profile generally shows a transition within a very small radial range ($\Delta r/a \ll 0.1$) having a large slope across the transitioning surface \cite{Toda2004,Toda2018} because the neoclassical hydrogen flux is overestimated in the limit $E_r\to 0$.
In the measurement of \cite{Ido2010}, however, the transition width appears to be $\Delta r/a=0.1 \sim 0.2$.
This size is of the same scale as the typical ion drift orbit width and thus qualitatively consistent with our present simulation result in which the FOW effect is included {\it per se}.
Therefore, if the transition of the $E_r$ sign is a fundamental property of impurity hole plasmas, a global code which essentially includes the FOW effect has an advantage over local codes.
Still, it has been shown that the local neoclassical models retaining the tangential magnetic drift can yield results close to those of the global calculations of FORTEC-3D  \cite{Matsuoka2015,Fujita2019,VelascoKNOSOS, Velasco2019}.
Compared with the relative largeness of the computational cost of global simulation to that of local simulation, the additional computational cost required to include the tangential magnetic drift in a local code is not significant. 
Thus, if the local codes retaining tangential magnetic drift can
provide reliable estimations of the quantities we are interested in, namely ambipolar $E_r$, $\Phi_1$, and the neoclassical fluxes in impurity hole plasmas except for the neighborhood of the surface where $E_r=0$, the modified local-models also become useful tools to investigate the impurity hole phenomenon.
In particular, the models can be used to perform parameter surveys such as to analyze the dependence of the phenomenon on the $n$-$T$ profiles or on the magnetic field configuration. 
To assess this point, we first need to understand how inclusion of the magnetic drift works precisely and to what extent the global effects are necessary by comparing simulation results of the global and local models with tangential magnetic drift. 

The discrepancy of the transitioning points between the numerical results and the experimental data is also to be examined.
The radial domain where $E_r<0$ is roughly $r/a < 0.25$ in the numerical results while $r/a < 0.55$ in the experimental data.
Since our plasma profiles are not the same as those of the experimentally studied plasma, the transitioning points need not coincide with each other exactly.
However, it is not clear if the carbon flux would still be outwardly directed when the negative-$E_r$ region is expanded in numerical simulation.

Disagreement in the ambipolar $E_r$ profiles suggests the possibility that we still lack some essential factors to determine the ambipolar condition. 
If so, the impact of NBI heating is one of the candidates for it is shown that an NBI can change the $E_r$ profile \cite{Nunami2017impacts, Spong2007}.
In fact, the impurity holes in LHD are usually formed after the tangential NBI is applied. 
It is also worthwhile examining the effect of NBI fast-ion anisotropic distribution on the $\Phi_1$ potential profile \cite{Yamaguchi2017} in the impurity hole plasmas. 
The effect of NBI fast-ion on neoclassical and turbulent transport has also been studied to explain the hollow impurity density profiles observed in tokamaks \cite{Manas2020}.
In order to verify these effects, our numerical model needs to be enabled to incorporate the impact of NBI.

To evaluate the ambipolar $E_r$ and $\Phi_1$ in the present study, the local approximation and adiabatic response to $\Phi_1$ have been adopted for electrons, since solving the electron distribution function as well as the ions distribution functions by the global drift-kinetic model is much too time-consuming.
Though the local approximation for electron neoclassical transport is thought to be more accurate than for ions, it has been demonstrated that the discrepancy between local and global electron fluxes $\Gamma_e$ in LHD can become considerable as the electron collisionality becomes lower \cite{Matsuoka2011}.
The finite magnetic drift changes the $E_r$ dependence of $\Gamma_e$ as well as the main ion flux $\Gamma_i$ in low-collisionality plasmas and will result in different $E_r$ profile from the simulation which uses local $\Gamma_e$. 
In a study using local neoclassical model, non-adiabatic, kinetic treatment of electrons in the evaluation of $\Phi_1$ potential has been shown to be more important in electron-root than in ion-root \cite{Garcia2018}.
Although our main focus is on the near-axis region where only ion-root is found, the study shows the contribution of the kinetic electrons to $\Phi_1$ also becomes large where the root-transitioning occurs and $-d \Phi_1/dr$ exhibits an appreciable value on the transitioning surface.
The absolute value of $-d \Phi_1/dr$ is not very large (at most, 1 keV/m), but the contribution cannot be ignored to determine the exact position where the total radial electric field changes its sign.
However, predictions of local models for low-collisionality plasmas can be inaccurate especially when $E_r$ is close or equal to zero, as has been mentioned throughout this article.
It is an open question
whether the tendency found in the local but kinetic electron simulation holds true in global simulation or not.

 \subsection{Impact of $\Phi_1$}

In the present study, we found that the impact of $\Phi_1$ on the ambipolar radial electric field and particle fluxes of light ion species (H$^{1+}$ and He$^{2+}$) are insignificant.
On the contrary, as was most remarkable for case B, it was shown that $\Phi_1$ does play a role in the transport of carbon impurities and can be non-negligible.
However, when the impurity flux is inwardly directed and its value is sufficiently large as that for case C, $\Phi_1$ contributes to driving the impurity flux further inwardly. 

This tendency of $\Phi_1$ to contribute to enhancing the absolute value of carbon flux and not to invert the sign of the flux was found in our previous study as well \cite{Fujita2020}.
That is, $\Phi_1$ drives the flux more outwardly if the flux is outwardly directed without $\Phi_1$ but it drives the flux more inwardly if the flux is inwardly directed without $\Phi_1$.
Whether $\Phi_1$ drives the flux inward or outward depends on the spectrum structures of the density variation and $\Phi_1$.
Even if the contribution of kinetic electrons turns out be non-negligible for some aspects under the plasma condition we have considered, it cannot be expected that the impact will be so large as to turn the spectrum structure of $\Phi_1$ determined by ions upside down.
Thus, $\Phi_1$ cannot solely fill the gap between the experimental observations and the conventional neoclassical analysis. 
This fact suggests that the global drift-kinetic model is one of the fundamental keys to explain the formation of an impurity hole, where the inward impurity turbulent flux should be balanced with the outward neoclassical flux, as we will discuss below.

 \subsection{Consistency between the particle fluxes and density profiles}\label{sec:disc:balance}

By comparing the results of cases A to C, we can infer the process of an impurity hole formation.
As long as the carbon density profile near the magnetic axis is not hollow as in case A or moderately hollow as in case B, the carbon flux near the axis can be outwardly directed.
Neglecting the impurity particle source and turbulent transport, our neoclassical simulation results then mean that carbon density near the axis continues to decrease and the density gradient becomes steeper accordingly. 
When the gradient becomes as steep as that for case C, the direction of the carbon flux inverts.
This suggests the process that the radial particle fluxes are balanced between the states corresponding to cases B and C, and the steady density profile is achieved there.

The density profile is determined by the particle balance equation
\begin{equation}
\label{eq:Gam_balance}
     \frac{\partial n_a}{\partial t}
     +\frac{1}{V'}\frac{d}{d r} (V' \Gamma_a^{\rm tot}) =S_a,
\end{equation}
where $S_a$ is a source term, $V'$ is the radial derivative of the volume enclosed by the flux surface $r$ and the total radial particle flux $\Gamma_a^{\rm tot}$ is given by the sum of the neoclassical contribution $\Gamma_a^{\rm NC}$ and the turbulent contribution $\Gamma_a^{\rm Trb}$:
\begin{equation}
      \Gamma_a^{\rm tot} =\Gamma_a^{\rm NC}+ \Gamma_a^{\rm Trb}.
\end{equation}
In the impurity hole discharge in LHD analyzed here, a particle source of carbon in the plasma core region existed only in an instant when a carbon pellet was injected \cite{Mikkelsen2014}. 
Figure 5 in \cite{Mikkelsen2014} also shows that the impurity hole is a transition phenomenon in which the hollow C$^{6+}$ density decays in time. The time scale of density decay is estimated as $\tau_{\rm  decay}=[\pd(\ln n_{C})/\pd t]^{-1}\sim O(0.1  {\rm s}$) from the figure. On the other hand, in the FORTEC-3D simulations, the ambipolar condition was achieved in $10\tau_c\sim 0.001$s where $\tau_c$  is the collision time of carbon, which is a much shorter time scale than $\tau_{\rm decay}$. 
Therefore, we can analyze the particle balance in the impurity hole plasma by approximating $\pd n_c/\pd t=S_c\simeq 0$ in eq.(\ref{eq:Gam_balance}).

For a steady state without a particle source, the neoclassical and the turbulent contributions must cancel each other so that the total flux vanishes. 
The particle flux we have investigated in this article is only the neoclassical part and we thus cannot make a rigorous argument about whether the obtained profiles of the carbon flux and the carbon density are consistent.
Nevertheless, an existing literature is in favor of our result. 
The study \cite{Nunami2020} has investigated both the local neoclassical and ITG-driven turbulent fluxes in impurity hole plasmas by using a nonlinear gyrokinetic simulation code GKV.
It is shown in the study that the value of the ambipolar $E_r$ should be positive at least around $0.52<r/a<0.61$ for both contributions of the fluxes to balance each other.

It was shown in the study that the carbon nonlinear turbulent flux on the surfaces at $r/a=0.52$ and $r/a=0.61$, where the hollow $n_C$ profile is formed, are always negative in spite of a wide range scan of local carbon density and temperature gradients around the nominal values of them in case A.
The inward pinch of impurity turbulent flux was also predicted in a quasilinear estimation previously \cite{Mikkelsen2014}.
In \cite{Nunami2020}, it was conjectured that $\Gamma_{\rm C}^{\rm NC}$ on the flux surface where the impurity hole is formed should be positive for some reason so as to balance with negative $\Gamma_{\rm C}^{\rm Trb}$. 
The authors of the study pointed out that $E_r$ and $\Gamma_c^{\rm NC}$ (from local model) can become positive by the effect of tangential NBI torque. 
In the present study, we found that the positive $E_r$ and $\Gamma_{\rm C}^{\rm NC}$ appear as solutions of global neoclassical transport simulation.

\begin{figure}
 \centerline{\includegraphics[scale=0.4, angle=-90]
  {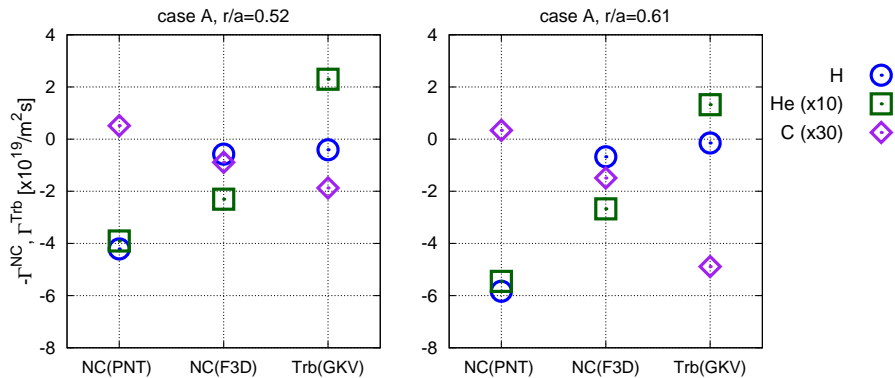}}
      \caption{Comparisons of $-\Gamma_a^{\rm NC}$ obtained with PENTA (left) and with FORTEC-3D (center), respectively and $\Gamma_a^{\rm Trb}$ obtained with GKV in \cite{Nunami2020} (right).
      Note that the values of the helium flux and the carbon flux are being multiplied by $10$ and $30$, respectively, for visualization purposes.
      The left figure shows the comparison on the flux surface at $r/a=0.52$ and the right figure at $r/a=0.61$, respectively.}
      \label{fig:NCvsTrb}
\end{figure}

Figure \ref{fig:NCvsTrb} compares the $-\Gamma_a^{\rm NC}$ calculated with PENTA and with FORTEC-3D in this work, respectively against $\Gamma_a^{\rm Trb}$ obtained with a local gyrokinetic code GKV in \cite{Nunami2020} at $r/a=0.52$ and $r/a=0.61$.
All the calculations are done using the same $n$-$T$ profiles which correspond to those for case A in this work.
Except for the FORTEC-3D calculation, $\Phi_1$ is not considered.
As shown in section \ref{sec:resultA}, PENTA finds an ion-root while FORTEC-3D finds an electron-root in the region considered here.
The gyrokinetic simulation assumes that the turbulent transport is independent of $E_r$.
The helium flux and the carbon flux are plotted being multiplied by $10$ and $30$, respectively, for visualization purposes
The first point to note is that, as has been discussed, the signs of the neoclassical carbon flux predicted by the PENTA (left column) and the FORTEC-3D (center column) simulations are opposite.
Only the global $-\Gamma_{\rm C}^{\rm NC}$ has the same sign as $\Gamma_{\rm C}^{\rm Trb}$ (right column) and they are only a factor $2$ to $3$ from being balanced.
On the other hand, both local and global simulations predict the opposite sign for $- \Gamma_{\rm He}^{\rm NC}$ from that of $\Gamma_{\rm He}^{\rm Trb}$ though the global result is closer to the turbulent contribution than the local result is. 
The largest discrepancy in the absolute value is seen in the hydrogen fluxes.
While the FORTEC-3D result is close to that of the GKV result, the PENTA result is far from balancing with the turbulent counterpart.
Therefore, in terms of the carbon and hydrogen flux balance, the scenario in which $E_r>0$ at least at $r/a>0.5$ is more plausible than the conventional ion-root scenario. 
The study \cite{Nunami2020} reaches the same conclusion by comparing the turbulent fluxes with the local neoclassical fluxes for an electron-root case as well as with the ion-root case considered here. 

It is to be noted that the turbulent fluxes show ``stiffness" and are much more susceptible to changes in the local density and temperature gradients
than neoclassical fluxes are.
In figure 4 of \cite{Nunami2020}, the turbulent electron and ion energy fluxes from gyrokinetic simulation were compared with their experimental observation values.
The simulation and observation values match if the $T_i$ gradient scale length is 
reduced by only about $30\%$ from the nominal value.
If we admit the ``flux-matching" condition \cite{Nunami2018} to explain the disagreement between observed and calculated energy fluxes by means of the ambiguity in the local temperature gradient, $\Gamma_{\rm C}^{\rm Trb}$ can then be reestimated by gyrokinetic simulation at the energy flux-matched temperature gradients of electrons and ions.
Indeed, it is confirmed that the energy flux-matched $\Gamma_{\rm C}^{\rm Trb}$
tends to be much in close balance with $\Gamma_{\rm C}^{\rm NC}$ obtained by the global
neoclassical simulation with $\Phi_1$ \cite{Nunami2020PSS,Nunami2020SPIG}.
In addition, with the modified temperature gradient, the turbulent helium flux is 
reduced by about one-half, while hydrogen flux remains almost unchanged.
Thus, considering the flux-matching of turbulent energy transport gives better particle balance $\Gamma_a^{\rm NC}+\Gamma_a^{\rm Trb}\simeq 0$ for each particle species.
Our result thus can be added to the agreement between the experimental and the numerical studies.
Nevertheless, further investigation on the flux balance using some sorts of integrated models of neoclassical and turbulent transport, especially for the near-axis region, is necessary to make a more precise argument.

In summary, by global simulation, we reproduced the ambipolar $E_r$ which changes its sign along the minor radius and the outward neoclassical carbon flux.
These are two aspects of impurity hole plasmas observed by experiments but have not been captured by local neoclassical simulations.
Also, the neoclassical and the turbulent particle fluxes nearly balance with each other which can explain the particle balance in the impurity hole plasma.

%% file: Body/Acknowledgement.tex
\section*{Acknowledgments} 

This work is performed under the auspices of the NIFS Collaborative Research Program, No. NIFS18KNST132. 
The simulations in this article were carried out on Plasma Simulator in NIFS, Japan and on JFRS-1 supercomputer system at Computational Simulation Centre of International Fusion Energy Research Centre (IFERC-CSC) in Rokkasho Fusion Institute of QST (Aomori, Japan).